\documentclass[10pt,aps,prl,twocolumn,showpacs,amsmath,amssymb,footinbib]{revtex4-1}
\usepackage{graphicx}% Include figure files
\usepackage{epstopdf,mathrsfs}
\usepackage{booktabs}
\usepackage{bbold}
\usepackage{epsfig}
\usepackage{dcolumn}% Align table columns on decimal point
\usepackage{bm}% bold math
\usepackage{braket}
\usepackage{CJK}
\begin{document}

%\preprint{APS/123-QED}

\title{Experimental observation of fractional topological phases with photonic qudits}
\author{A. A. Matoso\textsuperscript{1}}
\author{X. S\'{a}nchez-Lozano\textsuperscript{1,2}}
\author{W. M. Pimenta\textsuperscript{1,3}}
\author{P. Machado\textsuperscript{1}}
\author{B. Marques\textsuperscript{1,4}}
\author{F. Sciarrino\textsuperscript{5}}
\author{L. E. Oxman\textsuperscript{6}}
\author{A. Z. Khoury\textsuperscript{6}}
\author{S. P\'{a}dua\textsuperscript{1}}
\affiliation{
\textsuperscript{1}Departamento de F\'{\i}sica, Universidade Federal de Minas Gerais, 31270-901 Belo Horizonte, Minas Gerais, Brazil\\
\textsuperscript{2}Departamento de F\'{\i}sica - DCI, Universidad de Guanajuato P. O. Box E-143, 37150, Le\'{o}n, Gto., M\'{e}xico\\
\textsuperscript{3}Instituto de F\'{\i}sica, Universidad Aut\'{o}noma de San Luis Potos\'{\i}, San Luis Potos\'{\i} 78290, M\'{e}xico\\
\textsuperscript{4}Instituto de F\'{\i}sica, Universidade de S\~{a}o Paulo, 05508-090 S\~{a}o Paulo, Brazil\\
\textsuperscript{5}Dipartimento di F\'{\i}sica, Universit\`a di Roma La Sapienza, 00185, Roma, Italy\\
\textsuperscript{6}Instituto de F\'{\i}sica, Universidade Federal Fluminense, 24210-346 Niter\'{o}i, Rio de Janeiro, Brazil}

\date{\today}

\begin{abstract}
Geometrical and topological phases play a fundamental role in quantum theory. Geometric phases have been proposed as a tool for implementing unitary gates for quantum computation. A fractional topological phase has been recently discovered for bipartite systems. The dimension of the Hilbert space determines the topological phase of entangled qudits under local unitary operations. Here we investigate fractional topological phases acquired by photonic entangled qudits. Photon pairs prepared as spatial qudits are operated inside a Sagnac
interferometer and the two-photon interference pattern reveals the topological phase as fringes shifts when local operations are performed. Dimensions $d = 2, 3$ and $4$ were tested, showing the expected theoretical values.
\end{abstract}

%\pacs{42.50.Dv, 42.65.Lm, 3.65.Ud, 03.65.Vf, 03.67.Mn, 07.60.Ly}% PACS
% Classification Scheme.
%\keywords{??????}%Use showkeys class option if keyword
%display desired
\maketitle

Geometrical phases were introduced long ago in a seminal work on polarization transformations in classical optics \cite{1956Pancharatnam,1975Pancharatnam}. Later, geometrical and topological features were demonstrated to play a fundamental role in the realm of quantum theory \cite{1959Aharanov,1984Berry}. Tomita and Chiao verified experimentally the existence of Berry's phase and its topological properties at the classical level in helically wound optical fiber \cite{1986Tomita}. Geometric phases were measured by using coincidence detection of photon pairs produced in parametric down-conversion in conjunction with a Michelson interferometer \cite{1991Kwiat,1995Brendel}, by changing adiabatically the polarization state of the photon pairs \cite{1997Strekalov},  with single photons in a mixed state of polarization by using a Mach-Zenhder interferometer \cite{2005Ericsson} or in a polarization pure state by using a polarimetric technique \cite{2014Dezela}.   More recently, they have been proposed as a robust tool for implementing unitary gates for quantum computation \cite{2000Jones,2001Duan}. From this perspective, the geometric phase on entangled bipartite systems and the role of entanglement in its topological nature have been discussed both theoretically \cite{2000Sjoqvist,2003Milman,2006Milman} and experimentally \cite{2007Souza,2007Du,2014Loredo} for two-qubit systems. Later, they were generalized to pairs of qudits of any dimension \cite{2011Oxman,2014Khoury} and to multiple qubits \cite{2012Johansson}, showing that the dimension of the Hilbert space plays a crucial role in determining fractional topological phases in both cases. A recent review on this rich subject can be found in Ref.\cite{2015Sjoqvist}. Although experimental schemes for demonstrating these fractional values have been proposed \cite{2013Tutu,2013Johansson}, they have not been implemented so far. In a broader context, fractional phases may be revealed in quantum Hall systems, related to different homotopy classes in the configuration space of anyons. This nontrivial topology has been conjectured to be a possible resource for fault tolerant quantum computation \cite{2003Kitaev}.

In this work we present experimental results of fractional topological phase measurements on entangled qudits. A qudit is a quantum state that belongs to a $d$-dimensional Hilbert space ($d > 2$). A quantum system in a general qudit state can be written in terms of $d$ states that form a basis in the $d$-dimensional Hilbert space. Photonic qudits with dimensions $d=3$ and $4$ and qubits ($d = 2$) were encoded on the transverse positions of quantum correlated photon pairs generated by spontaneous parametric down conversion (SPDC). The photon pair is subjected to local unitary applied to their spatial degree of freedom using a spatial light modulator (SLM). In order to eliminate any dynamical phase contribution, these operations are restricted to SU($d$). The topological phases are measured as the phase shift of an interference pattern with respect to a reference pattern obtained without the SU($d$) operation. For SU($d$) transformations the dynamical phase vanishes identically. This has been demonstrated in three other previous theoretical works \cite{2011Oxman,2014Khoury,2013Johansson}. Fractional topological phases are observed through polarization-controlled two-photon interference \cite{2001Franca,2003Caetano}. Mukunda and Simon added an important contribution to the understanding of geometric phases in terms of a kinematic approach \cite{1993Mukunda}. When applied to a pair of $d$-dimensional entangled systems (qudits $A$ and $B$), following a cyclic evolution under local unitary operations, this approach allows the derivation of a general formula, $\phi_g = \frac{2n\pi}{d} - \sqrt{C_m^2 - C^2} \,(\Phi_A + \Phi_B)\quad(n\in\mathbb{N})$,
where $C$ is the I-concurrence, $C_m$ its maximum value ($C_m=\sqrt{2(d-1)/d}$) and $\Phi_{A(B)}$ is a phase contribution dependent on the whole evolution history of qudit $A(B)$\cite{2011Oxman,2014Khoury}. For example, for qubits, this phase contribution encompasses the usual Bloch sphere solid angle. Note that Bloch sphere representation is restricted to SU(2), so that it does not apply to higher dimensions ($d > 2$).For maximally entangled states $(C=C_m)$, only the fractional values $2n\pi/d$ can appear under local unitary evolutions. Our results show clear evidence of the $2\pi/d$ ($n=1$) theoretical prediction. The term fractional phases is used here in analogy to the phase acquired by the quantum state of identical particles when they are interchanged. In three dimensions we have $0$ for bosons and $\pi$ for fermions. However, the quantum state of identical particles in two dimensions can acquire fractional phases of the form $\frac{2\pi}{n}$ when the particles are interchanged. This is usually referred to as fractional statistics and the corresponding particles are called anyons. It has potential applications to quantum Hall effect and fault tolerant quantum computation as mentioned above. There is already a
quite extensive literature on these subjects \cite{1977Myrheim,1982Wilczek,1984Wu,1984Halperin,1984Wilczek,1985Zee}. However, the connection between fractional topological phases and high dimensional entangled states is recent and the results shown here represent an experimental demonstration.

\begin{figure}[htbp]
	\centering
		\includegraphics[height=5cm, width=6cm]{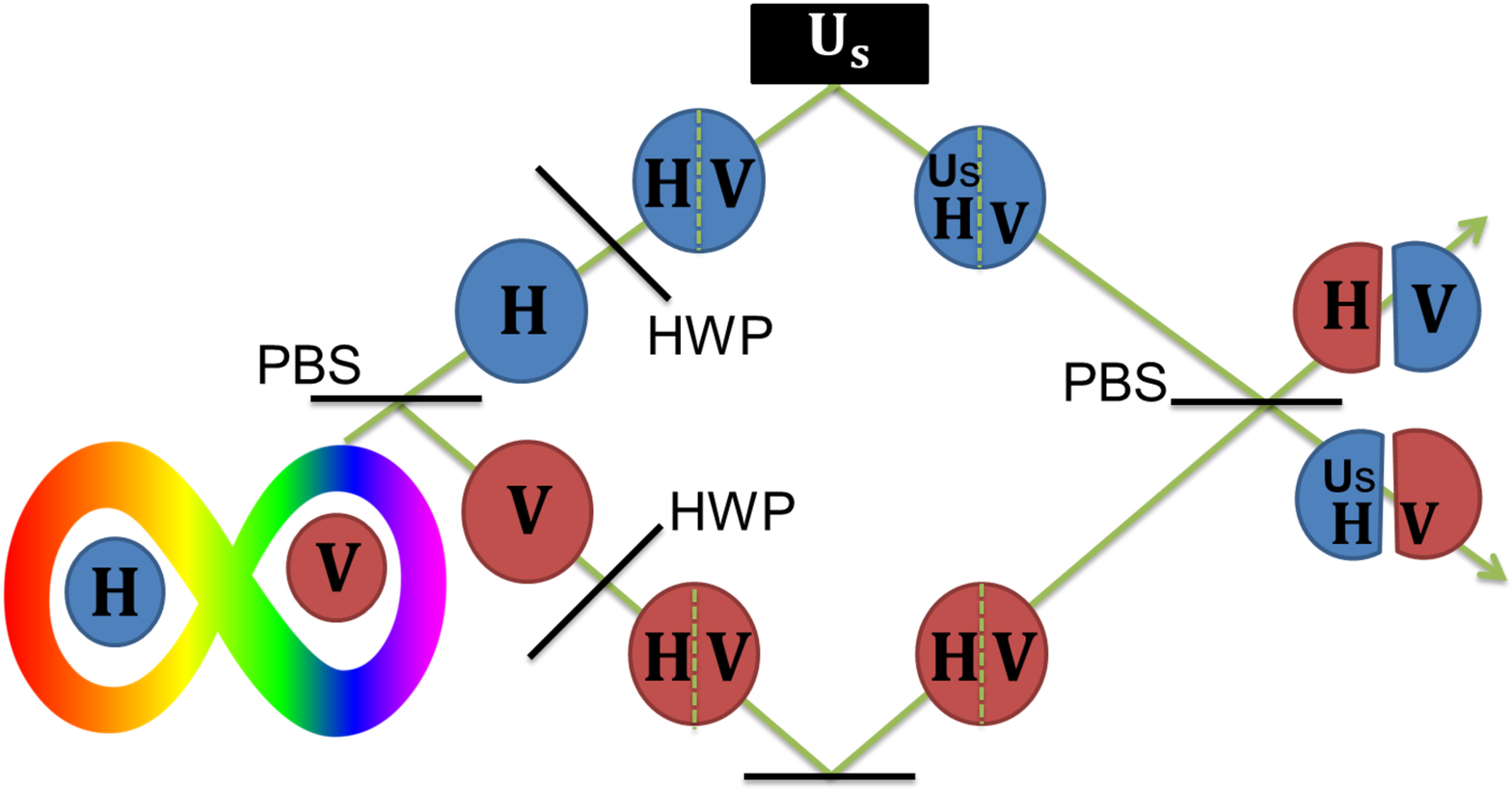}
	\caption[Experimental setup]{Scheme for measuring  the fractional topological phase of qudit state in the transversal path degree of freedom. Two photons entangled in their transversal paths with orthogonal polarization are split by a polarizing beam splitter (PBS) in two different longitudinal path. Each photon crossed a half wave plate (HWP) and propagate in a superposition of vertical and horizontal polarization states.
A local unitary operation $U_s$ is then applied to the transversal path degree of freedom of one of the photons conditioned to the horizontal polarization component. }
	\label{conceptual}
\end{figure}

Measurement of the topological phase for qudits is a challenging task. First we have to prepare a two-qudit photonic entangled state close to a maximum entangled state. Then, it is necessary to apply a suitable local unitary operation to one of the qudits of the pair. The third requirement is to set an interferometer with two possible arms where the two-qudit initial state interferes with the same state transformed by the SU($d$) operation. Fig. \ref{conceptual} shows schematically how we measured the topological phase. Two photons entangled in their transversal paths with orthogonal polarization are split by a polarizing beam splitter (PBS) in two different longitudinal paths. After each photon crossed a half wave plate (HWP) they propagate in a superposition of vertical and horizontal polarization states. A local unitary operation $U_s$ is then applied to the transversal path degree of freedom restricted to the horizontal component of one of the photons. Now the two photons exit the second PBS, have their polarizations rotated by 45° and are detected in coincidence after their horizontal components are filtered by a polarizer (not shown in Fig. \ref{conceptual}). We detect the interference between the entangled two-photon state in path variables and the same state after the application of a local unitary operation to the path degrees of freedom.

\begin{figure}[htbp]
	\centering
		\includegraphics[height=6.0cm, width=7.8cm]{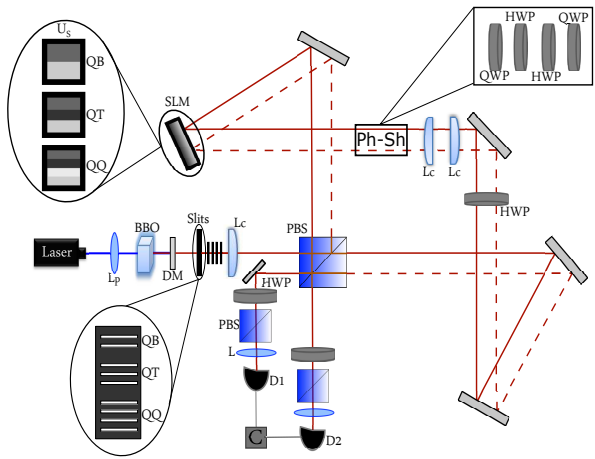}
	\caption[Experimental setup]{Experimental setup for measuring fractional topological phases:
a 355nm CW (continuous wave) laser beam is focused by a 30cm focal length lens (L$_p$) and pumps a 1mm $\beta$-barium borate (BBO)
crystal, generating collinear 710nm photon pairs in a type II phase matching configuration. A dichroic mirror (DM)
reflects the pump beam, while the photon pairs cross a multiple-slit placed at the focal plane of L$_p$.
The slits are 100$~\mu$m wide and their separation is 250$~\mu$m. A set of four thin ($\sim$1mm) quartz plates compensates the optical path difference between the orthogonally polarized photons exiting the BBO. Orthogonally polarized photons enter a Sagnac interferometer with a PBS as the input-output port. The SLM and the wave plates inside the interferometer allow the application of the local unitary transformations. The phase-shifter (Ph-Sh) is composed, from the right to the left, of a quarter wave plate (QWP) at 45°, a half wave plate (HWP) at $\phi$, another HWP at $\phi + \theta$, and another QWP at 45°. A 5cm focal length cylindrical lens (L$_c$) placed before the interferometer and two others in a telescope configuration inside it are used to image the slits on the SLM. Interference filters of 10nm bandwidth centered at 710nm are placed before the single photon detectors D$_1$ and D$_2$. Microscope lenses couple the photons into multimode fibers coupled to the detectors. Coincidence counts are registered in a time window of 5ns.}
	\label{setup}
\end{figure}

The qudits are encoded in the transverse paths of the photon pairs generated by SPDC and therefore the photon transverse paths must be well defined during their propagation inside the interferometer.
Polarization is used as an ancillary degree of freedom for conditional operation on the qudits. Fractional topological phases are obtained from the fringe displacement caused by local $SU(d)$ operations applied to the spatial qudits. The experimental setup is shown in Fig. \ref{setup}. A spherical lens L$_p\,$ placed before the crystal, focuses the pump laser beam on the multislit plane, dictating the spatial correlations of the down-converted photons \cite{2004Neves,2005Neves,2010Walborn}. The slits are placed after the crystal as shown in Fig. \ref{setup}. The number of slits set the dimension of the photon path states. A slide containing different multiple-slit arrays (lower inset in Fig. \ref{setup}) can be displaced vertically such that a two-qudit state with a different dimension can be prepared without modifying the setup.
Quartz plates introduced after the multiple-slit  correct the delay between the orthogonally polarized photons providing indistinguishable photon states in longitudinal space-time variables \cite{1987HOM,1994Shih,1997DeMartini}. The three cylindrical lenses L$_c$ in the setup are essential to keep the transverse paths well defined.

After the photon pair enters the interferometer, the two-photon state is written as

\begin{eqnarray}
\label{eq:psi0}
\ket{\Psi_0}=i\sum_{m,n=1}^d\alpha_{mn}\ket{mH}_s\ket{nV}_i,
\end{eqnarray}

where $\alpha_{mn}$ is the probability amplitude for the horizontally polarized photon (H) to cross slit $m$ and the vertically polarized phton (V) to cross slit $n\,$. By focusing the pump beam on the plane of the slits, we produce antisymmetric maximally entangled states \cite{2004Neves,2005Neves,2010Walborn}. For example, for $d=3$ the only amplitudes different from zero in Eq. \eqref{eq:psi0} are $\alpha_{13}=\alpha_{22}=\alpha_{31}=1/\sqrt{3}$ \cite{2004Neves}. The SLM in the Sagnac interferometer modulates only the phase of horizontally polarized photons, acting as a regular mirror for vertically polarized photons.

Signal and idler photons are labeled with horizontal and vertical polarizations, respectively. After crossing the PBS, the signal photon passes through a HWP that rotates its polarization by 45°, making $\ket{mH}_s\rightarrow(\ket{mH}_s+\ket{mV}_s)/\sqrt{2}$. Then, it follows through a phase shifter (Ph-Sh) composed by two quarter wave-plates (QWP) oriented at 45°, one HWP at a fixed angle $\phi$ and another HWP at a variable angle $\phi + \theta$ (upper right inset of Fig. \ref{setup}). The Ph-Sh introduces a $4 \theta$ phase difference between the horizontal and vertical polarization components \cite{1941Jones}, making $(\ket{mH}_s+\ket{mV}_s)\rightarrow(\ket{mH}_s + e^{4i\theta}\ket{mV}_s)$. At the SLM, an unitary operation $U_S = \sum_{k=1}^{d} e^{i\xi_k} \ket{k}\bra{k}$ is applied to the horizontal polarization component, making $(\ket{mH}_s + e^{4i\theta}\ket{mV}_s)\rightarrow(e^{i\xi_m}\ket{mH}_s + e^{4i\theta}\ket{mV}_s)$.
$\xi_k$ is the path phase added to the photons that cross the slit $k$. This operation is implemented by addressing $d$ rectangular windows on the SLM (upper left inset of Fig. \ref{setup}), each one matching a given slit image with a specific gray scale related to the corresponding value of $\xi_m$. The phase response of the reflecting SLM varies with the gray scale on the liquid crystal screen \cite{2005Goodman}. The discrete topological phases are expected for unitary transformations restricted to SU($d$), which in this case means $\sum_k \xi_k = 0$. The SLM is controlled to fulfill this condition.For example, for qutrits ($d = 3$) this operator is a diagonal 3 × 3 matrix [exp$i\phi_1$, exp$i\phi_2$, exp$i\phi_3$], and the SLM is programed to make $\phi_1 + \phi_2 + \phi_3 = 0$ such that the operator performed is SU(3) (unit determinant).

The idler photon has vertical polarization, being insensitive to the SLM. After the Ph-Sh and the HWP, the idler state components follow the sequence $\ket{nV}_i\rightarrow e^{2i\theta}\ket{nH}_i\rightarrow e^{2i\theta}(\ket{nH}_i-\ket{nV}_i)/\sqrt{2}$.
At the output of the Sagnac interferometer, the photons are recombined by the PBS in the two-qudit quantum state 
$\ket{\Psi_1} = \frac{1}{2}\sum_{m,n=1}^d \alpha_{mn} \left(e^{i\xi_{m}}\ket{mH}_s + e^{4i\theta}\ket{mH}_s\right) \otimes e^{2i\theta} \left(\ket{nH}_i - \ket{nV}_i \right)$
An irrelevant overall phase factor is acquired by the idler photon in the Ph-Sh. Note that only the $\ket{mH}_s\ket{nH}_i$ and $\ket{mV}_s\ket{nV}_i$ components will contribute to the coincidence counts, since in those cases the photons will exit in different ports. The state post selected by the coincidence detection is 
$ \ket{\Psi_2} = \frac{1}{\sqrt{2}}\sum_{m,n=1}^d \alpha_{mn} \left(e^{i\xi_{m}}\ket{mH}_s\ket{nH}_i - e^{4i\theta}\ket{mV}_s\ket{nV}_i\right).$
This state cannot exhibit interference because the spatially modulated and the nonmodulated components are identified by the photon's polarization. In order to erase the polarization information, a HWP and a PBS are placed before each detector. The HWPs rotate the polarization components by $\pm$45° and the PBSs project all components on the horizontal polarization before detection.

The coincidence count on detectors D$_1$ and D$_2$ is proportional to the correlation function
$C(\mathbf{r}_1,\mathbf{r}_2) = \langle E_1^-(\mathbf{r}_1)E_2^-(\mathbf{r}_2)E_1^+(\mathbf{r}_2)E_2^+(\mathbf{r}_1)\rangle$
integrated over the detectors areas, where $E_j^+$ ($E_j^-$) is the positive (negative) frequency component of the electric-field
operator on detector D$_j$ (j = 1,2). In terms of the field operators before the HWPs we have $E_1^+=\frac{1}{\sqrt{2}}(iE_{sV}^+ + E_{iH}^+)$, and $E_2^+=\frac{1}{\sqrt{2}}(E_{sH}^+ + iE_{iV}^+)$, where $E_{s\mu}^+$ ($E_{i\nu}^+$) is the positive frequency component of the signal (idler) vector field operator for different polarizations ($\mu,\nu=H,V$) \cite{2013Tutu}. Each polarization component is expanded in terms of the slit mode functions $\eta_p(\textbf{r})$ as $E_{s\mu}^+=\sum_p a_{p\mu}\eta_p(\mathbf{r_s}),\quad E_{i\nu}^+=\sum_q b_{q\nu}\eta_q(\mathbf{r_i}),$ where the annihilation operators $a_{p\mu}$ and $b_{q\nu}$ act on signal and idler Fock states, respectively, as
$a_{p\mu}b_{q\nu}\ket{m\sigma}_s\ket{n\epsilon}_i=\delta_{pm}\delta_{\mu\sigma}\delta_{qn}\delta_{\nu\epsilon}\ket{0}_s\ket{0}_i$.
Photons are guided by lenses and focused into multimode fibers coupled to D$_1$ and  D$_2$.
This corresponds to integrate the correlation function over the spatial variables. Using the orthonormality condition
of the slit modes, we obtain the normalized  function $C = \frac{1}{4}\sum_{m,n}|\alpha_{mn}\,e^{i\xi_m}-e^{4i\theta}\alpha_{nm}|^2$.

As the Ph-Sh's angle $\theta$ is varied, the coincidence counts show interference fringes that depend on the photon path state being prepared by the slits ($\alpha_{mn}$) and on the SLM operation ($\xi_m$). The discrete topological phases are manifested when $SU(d)$ operations are performed on maximally entangled states with anti-symmetric coefficients $\alpha_{m,d-m+1}=1/\sqrt{d}$ and $\alpha_{mn} = 0$ for $n\neq d-m+1$. The SLM was programed to assume different configurations parametrized by $t$, satisfying $\sum_k \xi_k(t) = 0$ and $\xi_k(0)=0$ $\forall$ $k$. With these settings, the coincidence function becomes

\begin{eqnarray}
C(\theta,t) &=& \frac{1}{d}\sum_{m=1}^d \sin^2\left[\frac{\xi_m(t) - 4\theta}{2}\right].
\label{interfinal}
\end{eqnarray}

For different SLM operations the interference pattern $C(\theta,t)$ displaces and changes the visibility. Fractional topological phases are observed as the phase shifts of the interference fringes when the maximal visibility is recovered.
We have tested the discrete phases for two entangled qubits ($d=2$), qutrits ($d=3$) and ququarts ($d=4$). Photon pairs exiting a double-slit are prepared in a spatial two-qubit state. The state prepared is characterized by the coefficients $\alpha_{12}=\alpha_{21}=1/\sqrt{2}$. Inside the Sagnac interferometer, the SLM is set for

\begin{eqnarray}
\xi_1(t) = -\xi_2(t) = \pi t \qquad (0\leq t\leq 1)\;.
\label{eq:csid2}
\end{eqnarray}

\begin{figure}[htbp]
	\centering
	\includegraphics[width=6cm]{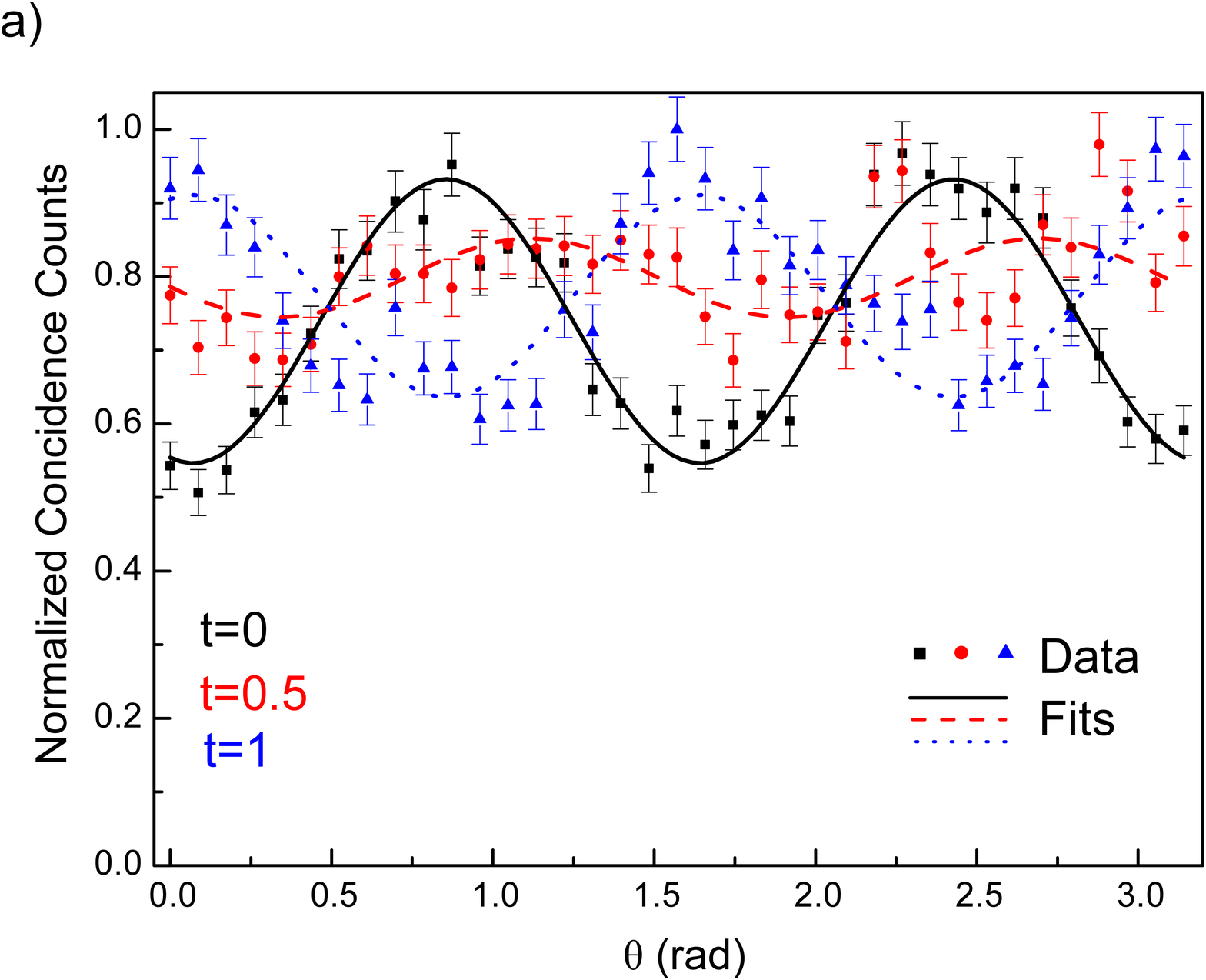}
	\includegraphics[width=6cm]{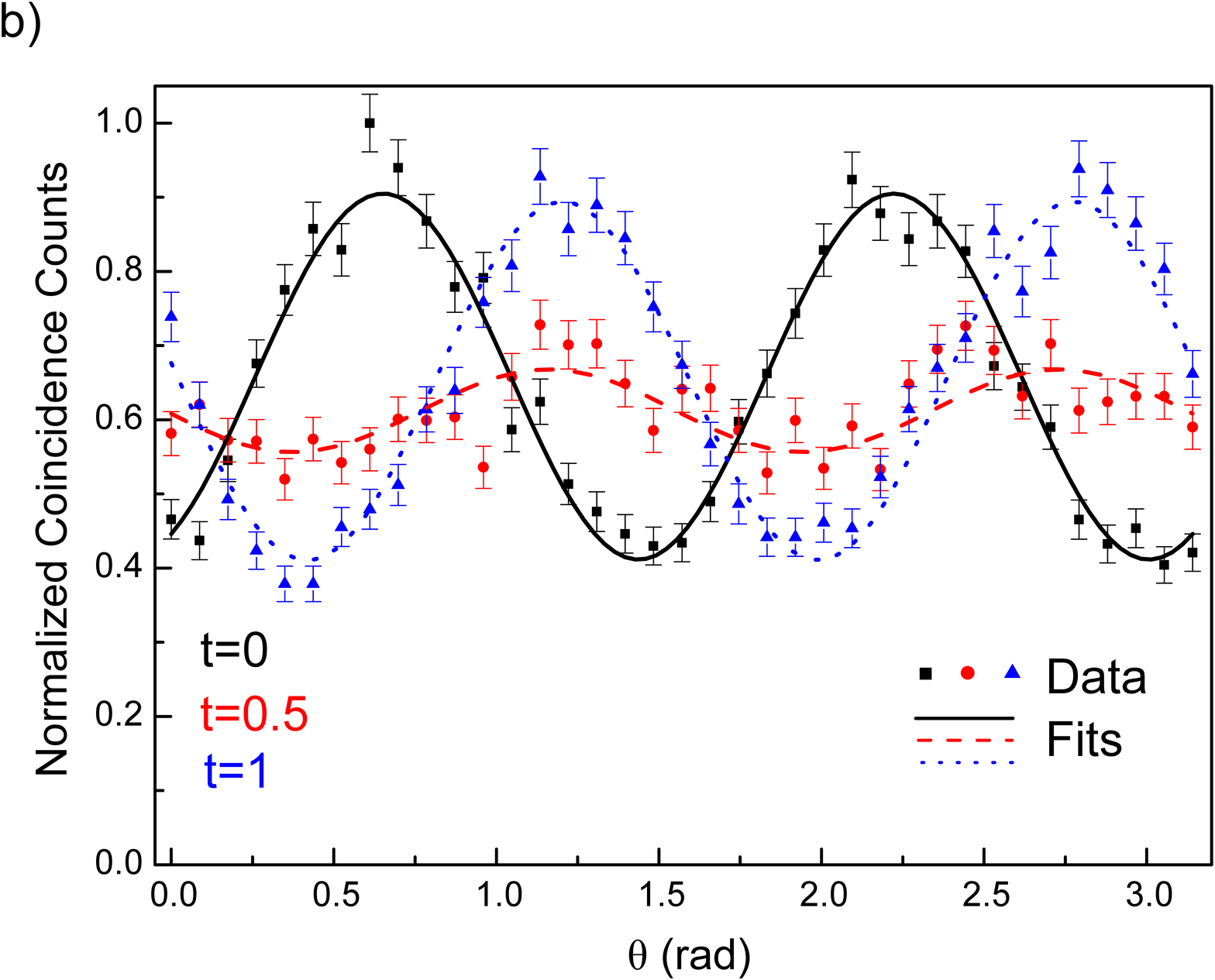}
	\includegraphics[width=6cm]{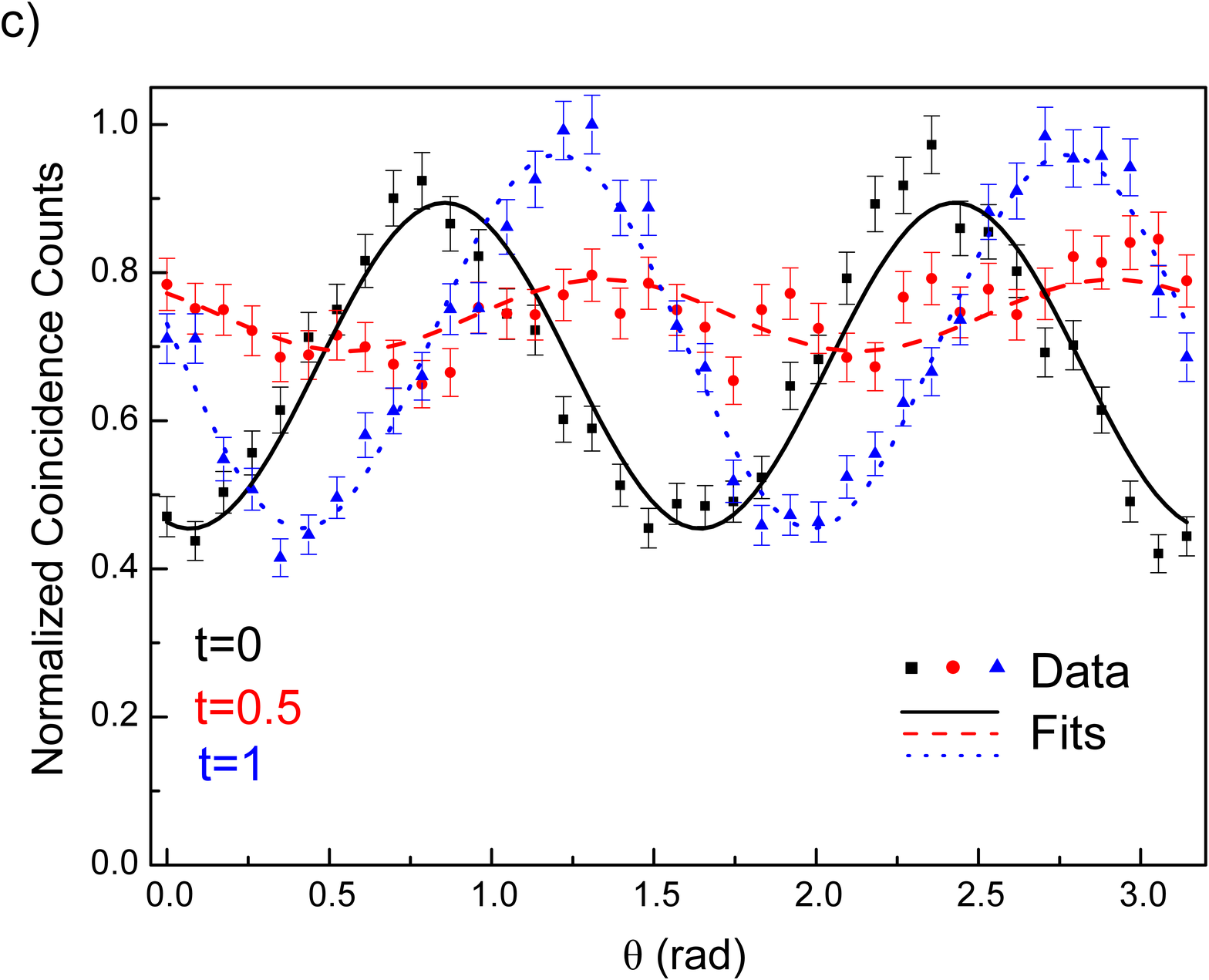}
	\caption[ex1]{Two photon interference fringes detected at the exit of the interferometer for a) qubits ($d=2$), b) qutrits ($d=3$) and c) ququarts ($d=4$). The experimental results are shown as dots with error bars and the numerical fits as lines. The results correspond to the unitary operations given by Eqs. \eqref{eq:csid2}-\eqref{unitary3} for three values of $t$: $t=0$ (black squares and black line), $t=0.5$ (red circles and red dashed line), and $t=1$ (blue triangles and blue dot line). The measured topological phases obtained from the experimental fits are: $(182 \pm 7)^{\circ}$ for the qubits, $(126 \pm 3)^{\circ}$ for qutrits and $(94 \pm 3)^{\circ}$ for ququarts.}
	\label{ex1}
\end{figure}

The control parameter $t$ is kept constant at each scan of the Ph-Sh's angle $\theta$. We use the parameter $t$ for condensing the
notation. It is interesting to visualize that the cyclic SU($d$) transformation implemented by the SLM is continuous and different unitary operations can be done in a cycle. The experimental results are shown in Fig. \ref{ex1}.a. The visibility of the two-photon interference curve decreases with $t$, reaching a minimum value at $t=0.5$, and returns to the original value when the full cycle is completed at $t=1$. The interference curve is then shifted by \textbf{$(182 \pm 7)^{\circ}$} which is confirmed by the theoretical value of $180^\circ$ expected for qubits. The experimental shift value was obtained from the fits shown in Fig. \ref{ex1}. The theoretical expression used for the experimental fit was $C(\theta) = \frac{1}{2}(1 − v \cos [a\theta + b])$, where $v$, $a$, and $b$ are fit parameters. In the second case, a 3-slit array prepares the two-qutrit entangled state \cite{wanderson1,wanderson2}.
The SLM was programed with

\begin{eqnarray}
\xi_1 &=& \frac{2\pi}{3}\left[2t-(2t-1)H(t-\frac{1}{2})\right]\;,
\nonumber\\
\xi_2 &=& -\frac{4\pi}{3}t \;, \xi_3 = \frac{2\pi}{3}(2t-1)H\left(t-\frac{1}{2}\right)\;
\label{unitary1lostvisib},
%\nonumber
\end{eqnarray}

with $t\in[0,1]$ and $H(t')$ is the Heaviside function. The resulting measurements for $t = 0, t = 0.5$ and $ t = 1$ are shown in Fig. \ref{ex1}.b. The reference curve was obtained with $t=0$. For $t=0.5$, the interference visibility decreases because the resulting evolution is not cyclic. A cyclic evolution is achieved for $t=1$, when the visibility recovers its initial value and the measured fringes are shifted by $(126 \pm 3)^{\circ}$. This value was obtained from the fits and is in good agreement with the theoretical prediction for qutrits ($\phi_{top}=2\pi/3$).

In the last example, a spatial two-ququart state is prepared. The SLM is programed to implement the following operation:

\begin{align}
\xi_1=\frac{\pi}{2}t, \xi_2=-\frac{\pi}{2}t+ \pi(1-2t)H\left(t-\frac{1}{2}\right), \label{unitary3} \\
\xi_3=\frac{3\pi}{2}t-\pi(1-2t)H\left(t-\frac{1}{2}\right), \xi_4=-\frac{3\pi}{2}t \;.\nonumber
\end{align}

\begin{figure}[htbp]
	\centering
	\includegraphics[width=6.5cm]{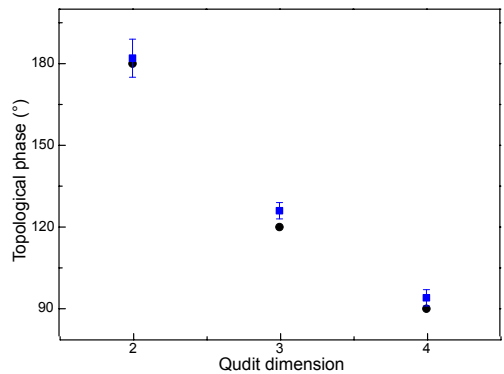}
	\caption[ex2]{Measured fractional topological phases (squares) and the expected theoretical values (balls) in terms of the dimension of the different prepared qudit states.}
	\label{ex2}
\end{figure}

The resulting interferences are shown in Fig. \ref{ex1}.c. We see that the visibility for $t=0.5$ is close to zero, and for $t=1$ the initial visibility is recovered. The observed phase shift was $(94 \pm 3)^{\circ}$ as expected for $d=4$. Single count oscillations
in all measurements are very small (less than 3\%) and are not responsible for the interference patterns measured by
the coincidences. Fig. \ref{ex2} shows the plot of the measured fractional topological phases and the expected theoretical values in terms of the dimension of the different qudit states.

Notice in Fig. \ref{ex1} that the visibilities of the measured interference patterns for $t=0$ and $t=1$ are lower than the predicted visibilities of $1.0$ \cite{2007Du}. These visibilities are independent of the dimension of the qudit states and are between $0.3$ and $0.4$. We checked for possible causes. Signal and idler temporal indistinguishability was tested by measuring a Hong-Ou-Mandel dip \cite{1987HOM,1994Shih} with photons exiting the crystal and after they exit the interferometer. Dips with visibilities $0.93 \pm 0.01$ and $0.88 \pm 0.03$  were measured by following the method described by G. Di Giuseppe et. al. in reference \cite{1997DeMartini} after the crystal and at the exit of the interferometer, respectively. Another cause is the mode mismatching between the transverse modes of photons exiting the slits and the coupled modes to the optical fibers that connect the interferometer exit and the detectors. This was tested with an attenuated He-Ne laser beam that crossed the interferometer and the measured visibility was $0.42 \pm 0.03$. The phase shifter was tested independently and oscillations with visibilities near $1.0$ were obtained. Therefore the main cause of the discrepancy between the predicted visibilities and the measured ones is the imperfect transverse and longitudinal mode matching. This discrepancy does not affect the measurement of the topological phase for qudits. The measured phase is obtained from the interference pattern displacement [SU($d$) operation applied; t = 1] when compared to a reference interference pattern [no SU($d$) operation applied; t = 0] as shown in Fig. \ref{ex1}. The interference pattern displacements are independent of the pattern visibilities.

In conclusion, we have measured the fractional topological phases acquired by entangled qudits following cyclic evolutions under local SU($d$) transformations. We have designed an experiment for measuring the topological phase acquired by qudits, when the photonic spatial structure is efficiently operated by a spatial light modulator. The results for maximally entangled states are conclusive. The experimental two-photon interference patterns confirm that under local SU($d$) operations, maximal visibility can only be attained with fractional phase shifts multiples of $2\pi/d$. The doors are now open to the experimental realization of qudit gates based on topological phases for implementing quantum algorithms.

This work was supported by CNPq, CAPES, FAPEMIG, National Institute of Science and Technology in Quantum Information, and the Science without Borders Program (Capes and CNPq, Brazil). X.S-L. acknowledges financial support from CONACyT-Mexico.
%\section*{References}

\end{document}